\documentclass[twocolumn,aps,prl,superscriptaddress]{revtex4}
\usepackage[utf8]{inputenc}
\usepackage{epsf,graphicx}
\usepackage{multirow}
\usepackage{amsmath,gensymb,amssymb}
\usepackage{lipsum}
\usepackage{dcolumn}
\usepackage{bm}
\usepackage{hyperref}
\usepackage{natbib}
\usepackage{color}
\usepackage{lipsum,graphicx}

\begin{document}

\title{Three-Dimensional Acoustic Turbulence: Weak Versus Strong}
\author{E.A. Kochurin}
\email{kochurin@iep.uran.ru}
\affiliation{Institute of Electrophysics, Ural Branch of RAS, 620016, Yekaterinburg,  Russia}
\affiliation{Skolkovo Institute of Science and Technology, 121205, Moscow, Russia}
\author{E.A. Kuznetsov}
\affiliation{Skolkovo Institute of Science and Technology, 121205, Moscow, Russia}
\affiliation{Lebedev Physical Institute, RAS, 119991, Moscow, Russia}
\affiliation{Landau Institute for Theoretical Physics, RAS, Chernogolovka, 142432, Moscow
region, Russia}
\date{\today}

\begin{abstract}
Direct numerical simulation of three-dimensional acoustic turbulence has
been performed for both weak and strong regimes. Within the weak turbulence,
we demonstrate the existence of the Zakharov-Sagdeev  spectrum
$\propto k^{-3/2}$
 not only for weak dispersion but in the
non-dispersion (ND) case as well. Such spectra in the
$k$-space are accompanied by jets in the form of narrow cones.
%For WD, they are concentrated  at
%the small $k$, but in the ND case, represent a set of well-separated jets for all $k$.
These distributions are realized due to small nonlinearity  compared with both dispersion/diffraction. Increasing pumping in the ND case due to dominant nonlinear effects  leads to the formation
of shocks. As a result, the acoustic turbulence turns into an ensemble of
random  shocks with the Kadomtsev-Petviashvili  spectrum.

\end{abstract}

\maketitle

\textbf{1.} As is known, the developed hydrodynamic turbulence at large
Reynolds numbers, $Re\gg 1$, in the inertial  interval represents an
example of a system with strong nonlinear interaction, when its energy
coincides with the interaction Hamiltonian. Another example relates to
acoustic turbulence which demonstrates both strong and weak regimes
depending on the ratio between nonlinearity and linear wave characteristics.
In this sense, acoustic turbulence is much more diverse and richer than
hydrodynamic turbulence. When nonlinear interaction of waves is small
compared to linear effects, we have a regime of weak turbulence \cite%
{KZ-book} which can be investigated perturbatively by using the random phase
approximation. Within the weak turbulence theory (WTT) for arbitrary wave
systems, ensembles of waves are described statistically in terms of the
corresponding kinetic equations \cite{KZ-book, naz-book}. This theory
assumes that each wave with its random phase moves long enough time almost
freely and undergoes very rarely small changes due to the nonlinear
interaction with other waves. To date, the WTT has been confirmed with high
accuracy for dispersive waves (e.g. %free-surface gravity and capillary waves
\cite{FM}). The situation changes for acoustic waves without dispersion
for which %are met in many physical systems (see, e.g. \cite{Gurevich, Galtier, KuznetsovMusher, grav, Shuryak}). In this case,
the resonant
conditions for three-wave interaction
\begin{equation}  \label{3wave}
\omega(\mathbf{k})=\omega(\mathbf{k_1})+ \omega(\mathbf{k_2}), \quad \mathbf{k%
}=\mathbf{k_1}+\mathbf{k_2}
\end{equation}
are satisfied only for collinear wave vectors $\mathbf{k_i}$, where $\omega(\mathbf{k})=kc_s$ is the linear dispersion relation and $c_s$ is the speed of sound.
These interacting waves thus form a ray in the $k$-space. Evidently,
translating to the system of coordinates moving with $c_s$ along the ray
makes this system strongly nonlinear. In 1D gas dynamics,
this nonlinearity leads to the breaking of acoustic waves in accordance with
the famous Riemann solution.

However, in a multi-dimensional situation by means of such transition it is
possible to exclude $c_{s}$ only for one given ray, for all other rays
propagating under some angles this exclusion does not work. If we take
continuously distributed rays with close propagation angles, we obtain an
acoustic beam, which, as is known, is subject to diffraction in the
transverse direction. As soon as a ray in the 3D case gets a transverse
width, let small, then such a ray will diffract. First time this effect was
discussed in the original paper by Zakharov and Sagdeev \cite{zs-70};
practically at the same time the analogous ideas were developed by Newell
and Aucoin \cite{Newell} (see, also \cite{AT}). From these arguments follows
that the weak turbulence regime for acoustic waves can be realized, as we
will show in this Letter, not only in the weak wave dispersion case but also
in the dispersionless situation due to  the diffraction.

For acoustic waves with weak positive wave dispersion,
\begin{equation}
\omega =kc_{s}(1+a^{2}k^{2}),\quad(a^{2}k^{2}\ll 1),  \label{disp1}
\end{equation}%
the resonant conditions (\ref{3wave}) are satisfied so that the interacting
waves instead of rays form cones with small angles $\sim (ak)^{2}$ (here $a$
is the dispersion length). The WTT, as known, is
applicable for small nonlinearity compared to weak dispersion. In this
mode, Zakharov and Sagdeev \cite{zakh65, zs-70} found exact isotropic
solution of the kinetic equation for 3D weak acoustic turbulence
%corresponding to the distribution
\begin{equation}
E(k)=C_{KZ}\left( \rho c_{s}\epsilon \right) ^{1/2}k^{-3/2},  \label{ZS}
\end{equation}%
the so-called Zakharov-Sagdeev (ZS) spectrum. Here $\epsilon $ is the energy
flux towards short wave scales, $C_{KZ}$ is the Kolmogorov-Zakharov (KZ)
constant, $\rho $ is density. The power dependence on $\epsilon $ in  (\ref{ZS}) with
the exponent $1/2$ corresponds to the resonant three-wave interactions (\ref%
{3wave}). In the long-wave limit such a spectrum is local and does not
depend on the wave dispersion length $a$ \cite{zakh65, zs-70}. However, the
2D WT spectrum contains the dependence on $a$ \cite{Naz22}.

With increasing pumping the nonlinear interaction becomes comparable to wave
dispersion that results in appearance of various coherent phenomena:
solitons, collapses, shock waves, etc. (see, e.g. \cite{ZakharovKuznetsov2012}).
%{KuznetsovTuritsyn1982, KuznetsovMusher}.
In this case the  system behavior is defined by interaction between
coherent structures and their incoherent partners.

In the dispersionless case, when nonlinear effects prevail over diffraction,
breaking of the acoustic waves leads to the formation of shocks. In this
regime, according to Kadomtsev and Petviashvili (KP) \cite{KP}, the acoustic
turbulence can be represented as an ensemble of randomly distributed shocks
that gives the KP spectrum
\begin{equation}
E(k)= \frac{2S_d c_s^2 \langle \delta \rho^2\rangle}{\pi \rho}k^{-2},
\label{KP}
\end{equation}
where $S_d$ is the number of shocks per unit length, $\langle \delta
\rho^2\rangle$ is the mean-square of density jumps at discontinuities (see,
e.g. \cite{Kuz04}). Notice  that exponent of the power-law
spectrum (\ref{KP}) is independent on the space dimension. In the 1D case,
this  is % nothing more than
the spectrum of Burgers turbulence \cite%
{Burgers}. Thus, there are two different approaches to the description of
acoustic non-dispersive turbulence predicting different spectra for the same
phenomenon.

The main goal of this Letter is to elucidate the reasons for the different
behavior of acoustic turbulence spectra by means of direct 3D numerical
simulations of the dynamical system. We will show that both spectra (\ref{ZS}%
) and (\ref{KP}) are actually realized, the transition between them being
controlled by the level of nonlinearity. Until now there has been no
convincing numerical evidence for the realization of both turbulent spectra
for 3D acoustic turbulence. Recently, for the weak wave dispersion we have
reported some preliminary results of direct simulations of the acoustic WT
\cite{KochurinKuznetsov}. In the given Letter, we present the new results
mainly concerning statistical properties of this turbulence. Note that, in the dispersionless case, up to now the WT regime has
not been numerically observed.

\textbf{2.} Direct numerical simulation of acoustic turbulence has been
carried out in the framework of the nonlinear string equation for a scalar
function $u(\mathbf{r},t)$ depending on three spatial coordinates $\mathbf{r}%
=\{x,y,z\}$ and time $t$ \cite{zakh65, zs-70}:
\begin{equation}
u_{tt}=\Delta u-2a^{2}\Delta ^{2}u+\Delta (u^{2}),  \label{eq0}
\end{equation}
where $a$ is the dispersion length, $\Delta$ is the Laplace operator. Here, we put $c_s=1$, and $\rho=1$. In the linear approximation, the
equation (\ref{eq0}) has the dispersion law
\begin{equation}
\omega ^{2}_k=k^{2}+2a^{2}k^{4},  \label{disp}
\end{equation}
which at $ka\ll 1$ transforms into (\ref%
{disp1}).

The equation (\ref{eq0}) can be represented as:
\begin{equation}
u_{t}=\frac{\delta H}{\delta \phi },\qquad \phi _{t}=-\frac{\delta H}{\delta
u},  \label{ham}
\end{equation}
where $u$ has the meaning of the density fluctuation and $\phi $ is of the
hydrodynamic potential (the velocity $\mathbf{v}=\nabla \phi$), and the
Hamiltonian $H$ is written as
\begin{equation*}
H=\frac{1}{2}\int \left[ \left( \nabla \phi \right) ^{2}+u^{2}\right] d%
\mathbf{r}+\int a^{2}(\nabla u)^{2}d\mathbf{r}+\frac{1}{3}\int u^{3}d\mathbf{%
r}\equiv
\end{equation*}
\begin{equation}
\equiv H_{1}+H_{2}+H_{3}.  \label{ham1}
\end{equation}
The first term $H_{1}$ is the sum of the kinetic and potential energies of
linear non-dispersion waves. The second term $H_{2}$ is responsible for the
wave dispersion, and $H_{3}$ describes the nonlinear interaction of waves.

Making the Fourier transform %, $f_{k}=(2\pi )^{-3/2}\int f(\mathbf{r})e^{-i%
%\mathbf{r}\cdot \mathbf{k}}d\mathbf{r}$,
and introducing the normal
variables $a_{k}$ and $a_{k}^{\ast }$,
\[
u_{k} =\left( \frac{k^{2}}{2\omega _{k}}\right) ^{1/2}(a_{k}+a_{-k}^{\ast
}), \, \phi _{k} =-i\left( \frac{\omega _{k}}{2k^{2}}\right)
^{1/2}(a_{k}-a_{-k}^{\ast }),
\]%
equations (\ref{ham}) take the standard form \cite{KZ-book, naz-book}:
\begin{equation}
\frac{\partial a_{k}}{\partial t}=-i\frac{\delta H}{\delta a_{k}^{\ast }},
\label{normal}
\end{equation}%
where
\begin{equation*}
H=\int \omega _{k}|a_{k}|^{2}d\mathbf{k}+\int V_{k_{1}k_{2}k_{3}}\left(
a_{k_{1}}^{\ast }a_{k_{2}}a_{k_{3}}+\right.
\end{equation*}%
\begin{equation*}
\left. +a_{k_{1}}a_{k_{2}}^{\ast }a_{k_{3}}^{\ast }\right) \delta \left(
\mathbf{k}_{1}-\mathbf{k}_{2}-\mathbf{k}_{3}\right) d\mathbf{k}_{1}d\mathbf{k%
}_{2}d\mathbf{k}_{3}.
\end{equation*}%
In the nonlinear Hamiltonian $H_{3}$, we left only one resonance term
corresponding to decay processes (\ref{3wave}). We will take into account  the dispersion only in the quadratic part of $H$
and ignore it in the
matrix element
\begin{equation}  \label{matrix}
V_{k_{1}k_{2}k_{3}}=\frac{1}{8\pi ^{3/2}}\left( k_{1}k_{2}k_{3}\right)
^{1/2}.
\end{equation}%
This expression coincides, up to a constant, with that for acoustic waves \cite{KZ-book,
zakh65}. Hence, the kinetic equation for the pair correlator $n_{k}$ ($%
\langle a_{k}^{\ast }a_{k_{1}}\rangle =n_{k}\delta (\mathbf{k}-\mathbf{k}%
_{1})$) in the WT approximation is written as
\begin{equation}
\frac{\partial n_{k}}{\partial t}=2\pi \int d\mathbf{k}_{1}d\mathbf{k}%
_{2}\left( T_{kk_{1}k_{2}}-T_{k_{1}kk_{2}}-T_{k_{2}kk_{1}}\right) ,
\label{kin}
\end{equation}%
where
\begin{equation*}
T_{kk_{1}k_{2}}=|V_{kk_{1}k_{2}}|^{2}(n_{k_{1}}n_{k_{2}}-n_{k}n_{k_{2}}
\end{equation*}%
\begin{equation*}
-n_{k}n_{k_{1}})\delta \left( \mathbf{k}-\mathbf{k}_{1}-\mathbf{k}%
_{2}\right) \delta \left( \omega _{k}-\omega _{k_{1}}-\omega _{k_{2}}\right)
.
\end{equation*}%
The turbulence spectrum $E(k)$ is found from the solution of this equation
after angle averaging (see \cite{SP, landau})
\begin{equation*}
E(k)=\frac{1}{(2\pi )^{3}}k^{2}\omega _{k}\int n(\mathbf{k})d\Omega .
\end{equation*}

As was first noted by Zakharov \cite{zakh65}, in the kinetic equation (\ref%
{kin}) in the isotropic case, the dispersion in $\omega _{k}$ can be
neglected, despite of a singularity due to the product of two $\delta $%
-functions which after angle averaging turns out to be integrable. As a
result, the kinetic equation admits, except thermodynamic distribution, a
stationary power-law solution of the Kolmogorov type: $n_{k}\propto k^{-9/2}$%
, which corresponds to the ZS spectrum (\ref{ZS}). The KZ
constant within (\ref{eq0}) for isotropic turbulence (see Supp. Mat. I  \cite{SP})
\begin{equation}
C_{KZ}\approx 0.22.  \label{KZaa}
\end{equation}%
The existence of the ZS spectrum in the inertial interval as a spectrum of
the Kolmogorov type was numerically confirmed in a number of papers (see, e.g. \cite%
{kin1,kin2} by  solving the kinetic equation (\ref{kin})
%. These  numerical studies of weak acoustic turbulence
%in the framework of (\ref{kin}) was carried out
only for isotropic distributions. In this Letter, we will show that in direct
numerical simulation of 3D acoustic turbulence described by
Eq. (\ref{kin}) supplemented by damping at high $k$ and pumping in the
region of long waves, the structure of the spectra is not isotropic
%especially
in the region of small $k$.

To model turbulence we introduce into the equations (\ref{ham}) both pumping
and damping terms:
\begin{equation}
u_{t}=-\Delta \phi +\mathcal{F}(\mathbf{k},t)-\gamma _{k}u,  \label{eq1}
\end{equation}
\begin{equation}
\phi _{t}=-u+2a^{2} \Delta u-u^{2},  \label{eq2}
\end{equation}
where the positive definite operator $\gamma _{k}$ responsible for
dissipation takes non-zeroth values at $k\geq k_d$ and the random forcing
term $\mathcal{F}(\mathbf{k},t)$ is narrowly localized in the region of
large scales $k\leq k_2$ with a maximum $F_0$ reached at $k_1$ ($k_d\gg k_2$%
). The instantaneous energy dissipation flux $\epsilon_N(t)$ is given as
\begin{equation}  \label{Ds}
\epsilon_N(t)=L^{-3}\int\limits_{L^3} u \hat {f}^{-1}(\gamma_k u_k) d\mathbf{%
r},
\end{equation}
where $\hat{f}^{-1}$ is the inverse Fourier transform, and $L$ is the system
size. In the quasi-stationary state, the energy flux is defined as a mean
value $\epsilon=\langle\epsilon_N\rangle=T^{-1}\int_0^T \epsilon_N(t) dt$.
The numerical simulation of the Eqs. (\ref{eq1}) and (\ref{eq2}) is carried
out with using accurate pseudo-spectral methods with the total number of
Fourier harmonics $N^3=512^3$ in a periodic box of size $L^3=(2\pi)^3$. We
set the parameters as follows: $k_1=3$, $k_2=6$, $k_d=110$. The details of
numerical methods are presented in  Supp. Mat. II \cite{SP}.

\begin{figure*}[t!]
	\centering
	\includegraphics[width=5.75in]{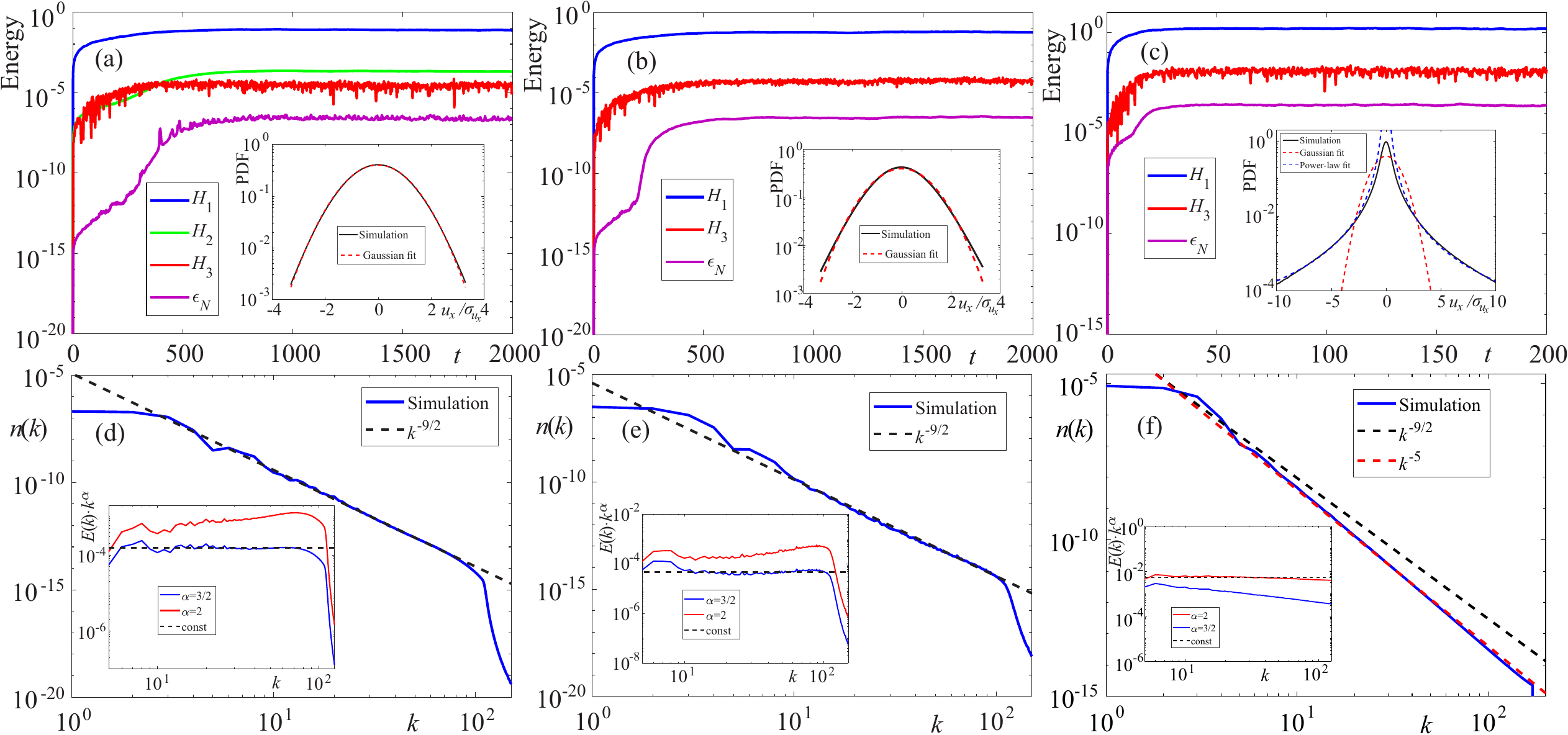}
	\caption{Energy contributions (\protect\ref{ham1}) and dissipation flux (%
		\protect\ref{Ds}) as functions of time for $a=2.5\cdot 10^{-3}$ (a) and for
		$a=0$ (b) and (c). The insets show PDF for  $u_{x}$ rescaled by
		its standard deviation $\protect\sigma _{u_{x}}$. Red dashed lines
		correspond to a Gaussian fit and blue dashed line is the power-law fit $%
		|u_{x}/\protect\sigma _{u_{x}}|^{3}$. The spatial spectra of wave action $n(k)$ are shown in
the quasi-stationary state for $%
		a=2.5\cdot 10^{-3}$ (a) and for $a=0$ (b) and (c). The insets are the
		compensated  turbulence energy spectra.}
	\label{fig1}
\end{figure*}

\textbf{3.} First, we consider the weakly dispersive regime with $a=2.5\cdot
10^{-3}$. Fig.~\ref{fig1}~(a) shows the temporal evolution of energy
contributions (\ref{ham1}). It can be seen the rather quick transition at $%
t\approx 500$ to the quasi-stationary chaotic regime. Both the dispersive
part $H_{2}$ and the nonlinear interaction energy $H_{3}$ turn out to be
small compared to $H_{1}$. The term $H_2$ exceeds $H_3$ by almost one order
of magnitude, which indicates the realization of a weakly nonlinear regime.
The inset to Fig.~\ref{fig1}~(a) shows the probability density function
(PDF) measured in the quasi-stationary state for the derivative $u_x $. The
PDF is close to a normal Gaussian distribution: the found Skewness $%
S=\langle u_x^3\rangle/\langle u_x^2\rangle^{3/2}$ is near $6.4\cdot 10^{-3}$%
, and Kurtosis $K=\langle u_x^4\rangle/\langle u_x^2\rangle^{2}\approx 3.31$%
. Deviations of these values from $S=0$ and $K=3 $ for Gaussian distribution
characterize extreme events. The wave action $n(k)$ averaging over angles
shown in Fig.~\ref{fig1}~(d) acquires a power-law behavior corresponding to
the ZS spectrum with high accuracy (see also the inset for the compensated
energy turbulence spectrum). The numerically found KZ constant $C_{KZ}\approx0.24$ is close to the theoretical prediction (\ref{KZaa}) (for more details Sec. III of Sup. Mat. \cite{SP}). To our knowledge, this is the first direct calculation of the KZ constant
for 3D acoustic turbulence.

\textbf{4.} The second numerical experiment is aimed to verify the possible
realization of WT regime in the absence of dispersion. Fig.~\ref{fig1}~(b)
shows that the system reaches the quasi-stationary state approximately an
the same time ($t\approx500$) as in the weakly dispersive case. The energy
pumping intensity $F_0$ was the same for both experiments, $F_0=1.25\cdot
10^{-3}$. The nonlinear interaction term $H_3$ turns out to be in three
orders of magnitude less than $H_1$. The PDF shown in the inset to Fig.~\ref%
{fig1}~(b) is close to Gaussian distribution with Skewness $S\approx4.9\cdot
10^{-2}$ and Kurtosis $K\approx3.47$. Fig.~(\ref{fig1})~(e) shows the
turbulence spectrum for non-dispersive regime $a=0$: $n_k$ spectrum turns
out in very well agreement with the power-law distribution, $\propto
k^{-9/2} $. The estimation for KZ constant gives $C_{KZ}\approx0.1$, i.e.
still has the same order with KZ constant (\ref{KZaa}).

Thus, upon the transition to a purely non-dispersive regime, the system
demonstrates weakly turbulent behavior with the ZS spectrum (\ref{ZS}). A
fundamental question arises: what is the mechanism for the generation of
weak acoustic turbulence in the absence of wave dispersion? Some differences
between two regimes are seen in Fig.~\ref{fig2}~(a) and (e): the
distribution of $u(\mathbf{r})$ in the weak dispersion (WD) case is smoother
compared to that of $a=0$. For the Fourier spectra $|u_k|$ in $k_z=0$ plane
(Fig.~\ref{fig2}~(b) and (f)), the difference becomes more observable. First,
for the WD case narrow jets form in the region of small $k$ (\cite{KochurinKuznetsov}). These jets widen with increasing $k$ due to wave
dispersion, and respectively the distribution becomes more isotopic. Unlike
WD, at $a=0$ the spectrum $|u_k|$ consists of a discrete set of narrow jets
of the conic type which do not intersect each other (see Fig.~\ref%
{fig2}~(c) and (g)). The corresponding isosurfaces of Fourier spectra $%
|u_k| $ shown in the Fig.~\ref{fig2}~(c) and (g) have the hedgehog form for
which observed rays (needles) appear as a result of three-wave nonlinear
interaction (\ref{3wave}).

\begin{figure*}[t!]
\centering
\includegraphics[width=6in]{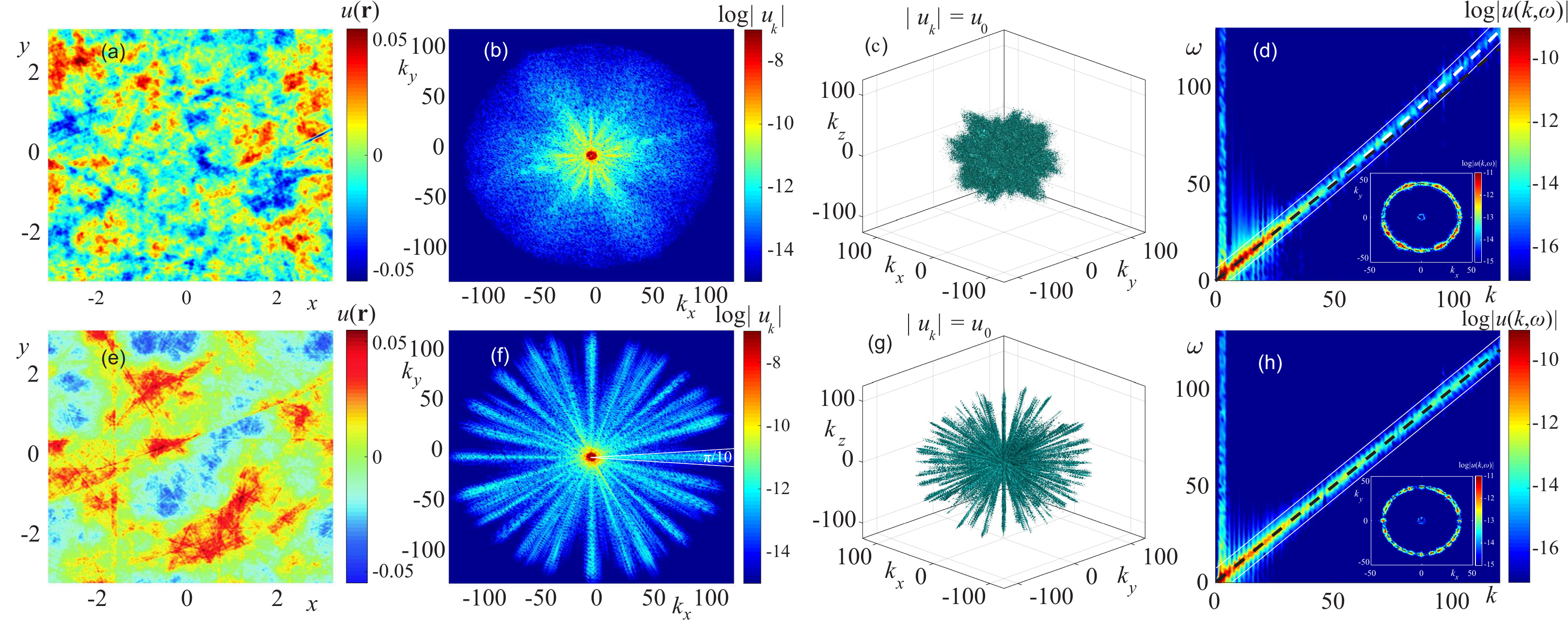}
\caption{ (a),(e): Dependence $|u(x,y)|$ at $z = 0$ - plane  at $t =
2000$; (b), (f): Fourier spectra of $|u(\mathbf{k})|$ as function of $k_x$
and $k_y$ at $k_z=0$ plane at $t = 2000$; (c),(g): Isosurfaces of the
Fourier spectra $|u_k|=5\cdot10^{-6}$ (in the 'hedgehog' form); (d), (h):
The space-time Fourier transform $|u(\mathbf{k},\protect\omega)|$, black and
white dashed lines correspond to the non-dispersive (\protect\ref{disp1})
and WD (\protect\ref{disp}) wave propagation, respectively,
white solid lines show the nonlinear frequency broadening $\protect\delta_{%
\protect\omega}$. The insets show $|u(\mathbf{k},\protect\omega_0)|$ with $%
\protect\omega_0=40$ on $k_z=0$ plane; (a)-(d) and (e)-(h) correspond to $%
a=2.5\cdot 10^{-3}$ and $a=0$, respectively.}
\label{fig2}
\end{figure*}

\begin{figure*}[t!]
	\centering
	\includegraphics[width=6in]{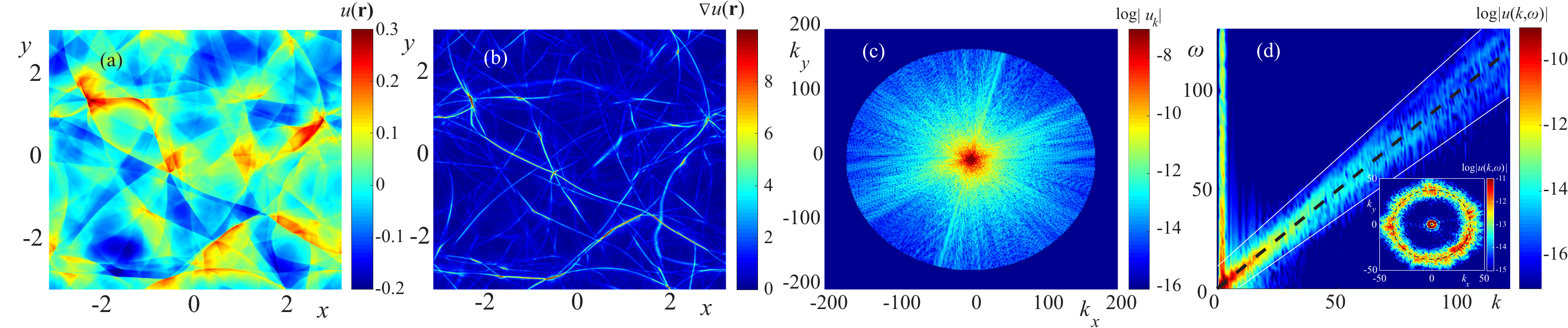}
	\caption{ (a), (b): Dependences $|u(\mathbf{r})|$ and $|\protect\nabla u (%
		\mathbf{r})|$ on the plane $z = 0$ for shock-wave regime, $t = 200$. (c):
		Fourier spectrum $|u_k|$ in the $k_z=0 $ plane is shown at $t = 200$. (d):
		The space-time Fourier transform $|u(\mathbf{k},\protect\omega)|$, the black
		dashed line corresponds to the non-dispersive wave propagation, (\protect\ref%
		{disp1}). White solid lines demonstrate the nonlinear frequency broadening $%
		\protect\delta_{\protect\omega}$. The inset shows $|u(\mathbf{k},\protect%
		\omega_0)|$ with $\protect\omega_0=40$ in $k_z=0$ section.}
	\label{fig3}
\end{figure*}

\textbf{5.} To clarify the WT mechanism in the non-dispersive (ND) regime, we
separate a jet directed along the $x$ axis by means of filtering (for
details, see   Sup. Mat. IV \cite{SP}). For such (narrow) jet , the
dispersion relation (\ref{disp1}) ($a=0$) can be expanded into a series
relatively the small parameter $\Omega _{0}=(k_{\perp}/k_x)$ (angle at the apex of a cone
directed along the $x$ axis)
\[
\omega _{k}=|\mathbf{k}|\approx k_{x}\left( 1+\Omega _{0}^2/2\right),
\]
so that $H_{1}$ represents a sum of two terms $H_{1}=\int
\omega _{k}n(k)d\mathbf{k}\approx \int {[k_{x}+\Omega _{0}^{2}k_{x}/2]n(k)d\mathbf{k}}=H_{\parallel }+H_{\perp }$, where $H_{\parallel }$ corresponds to the energy of acoustic beam propagating along $x$ direction and $H_{\perp
}$ defines its diffraction energy. To calculate these contributions, we
filtered the function $u(\mathbf{k})$ in a narrow cone
along $k_{x}$. Note that one selected cone (jet) occupies approximately
angle about $\pi /10$, see Fig.~\ref{fig2}~(f). Thus, the energy
contributions have a ratio $H_{\perp }/H_{\parallel }\approx 0.05$. Direct
calculation gives $H_{\perp }\approx 4.3\cdot 10^{-5}$, whereas the
nonlinear interaction energy for the selected jet is estimated as $%
H_{3}\approx 3.7\cdot 10^{-7}$. Thus, for the given jet the diffraction
energy is in two orders of magnitude larger nonlinear interaction energy
that explains appearance of the weak acoustic turbulence and the ZS spectrum (%
\ref{ZS}). The space-time Fourier transforms $|u(\mathbf{k}%
,\omega )|$ shown in the Fig.~\ref{fig2}~(d) and~(h) also indicate the
weakly nonlinear evolution. The figure does not reveal the
influence of any coherent and strongly nonlinear structures. Only a slight
frequency broadening $\delta _{\omega }$ %due to nonlinear effects
is observed.

Thus, weak acoustic turbulence can be realized in both weak dispersion and
non-dispersive cases. The criterion of WT in both regimes in some sense is
the same: the dispersion and/or diffraction effects should prevail
nonlinearity, which for acoustic waves is responsible for their breaking.

\textbf{6.} In the third numerical simulation, we investigate strong
acoustic turbulence in the non-dispersive case. To demonstrate such regime we
consider a situation of extremely large pumping amplitude when $F_{0}$ was
increased to $40\cdot 10^{-3}$, i.e., by more than an
order of magnitude larger in comparison with two previous numerical
experiments. Secondly, to suppress the \textquotedblleft
bottleneck\textquotedblright\ effect we set $k_{d}=1$, that provides the
energy dissipation in all ranges of $k$ (see Sup. Mat. V \cite{SP}).

For such conditions we observed a very rapid approaching to the stationary
state, $t\approx 50$, see Fig.~\ref{fig1}~(c). In the inset, one can see that the PDF is very different from the Gaussian distribution with highly
elongated non-Gaussian tails indicating the presence of extreme events.
Kurtosis reaches a very large value $K\approx46.37$. Skewness is estimated
as $S\approx 0.28$. Thus, the PDF demonstrates a high level of intermittency
in the system. The turbulence spectrum for this strongly nonlinear regime is
shown in Fig.~\ref{fig1}~(f): the exponent of $n(k)$ is very close to $-5$
corresponding to KP spectrum (\ref{KP}),  the energy spectrum $E(k)\sim k^{-2}$  is shown at the inset.

Fig.~\ref{fig3}~(a) and (b) show the spatial distributions of $u(\mathbf{r})$ and its
gradient $\nabla u (\mathbf{r})$ at the $z=0$ plane which demonstrate the
presence of a set of shock waves propagating under various angles.
The spatial turbulent spectrum shown in Fig.~\ref{fig3}~(c) clearly indicates
the quasi-isotropic behavior of wave energy in Fourier space.
Thus, with increasing the nonlinearity level, we observe a transition to
the state with strong intermittency. Fig.~\ref{fig3}~(d) shows the space-time
Fourier spectrum of $u(\mathbf{r},t)$ that allows us to estimate the
frequency broadening $\delta_{\omega }=1/\tau _{NL}$, where  $\tau _{NL}$ is characteristic nonlinear time. We have computed the parameter $\tau_{L}/\tau_{NL}$ with characteristic linear diffraction time $\tau _{L}=[\Omega_{0}^{2}k/2]^{-1}$ for both weak and strong WT regimes, see Sup. Mat. VI \cite{SP}. Comparison of these times shows that in the strong
turbulence regime, $\tau_{L}/\tau_{NL}$ is about 5. The latter means that the
wave breaking is more rapid process leading to formation of shocks randomly
distributed due to the chaotic pumping. This is a reason for the appearance of the KP spectrum ~\cite{KP}.

\textbf{7.} The main result of the work is that the ZS spectrum of 3D weak
acoustic turbulence (\ref{ZS}) can indeed be realized in both %
WD and ND regimes. The mechanism for the development of
WT is the divergence (diffraction) of acoustic waves preventing
their breaking. However, under action sufficiently large pumping,
acoustic turbulence turns into an ensemble of random shocks described by the
KP spectrum (\ref{KP}). Thus, the spectra (\ref{ZS}) and (\ref{KP})
correspond to different limiting cases of the weakly and strongly nonlinear
regimes of 3D acoustic turbulence.

The authors are grateful to V.E. Zakharov for very fruitful discussions at
the earlier stage of performance of this work. We also thank P.-L.
Sulem for one valuable remark. This work was supported by
the Russian Science Foundation (grant no. 19-72-30028).

\newpage

\onecolumngrid 

\section{Supplemental material to ``Three-Dimensional Acoustic Turbulence: Weak Versus Strong''}

In this Supplemental Material, we present additional data analysis related
to theoretical and numerical study of the three-dimensional acoustic
turbulence: weak turbulence theory for 3D acoustic waves (Sec. I), details of the
numerical model (Sec. II), estimations for the Kolmogorov-Zakharov
constants for both weakly dispersive and pure non-dispersive regimes (Sec.
III), analysis of a separate jet of the function $u(\mathbf{k})$ in Fourier
space (Sec. IV),  details for the suppression of ``bottleneck'' effect
arising for high level of nonlinearity (Sec. V), and characteristic nonlinear time comparison for weakly and strongly turbulent regimes of motion (Sec. VI).

\section{Weak turbulence theory for 3D acoustic waves}

Consider the wave kinetic equation (11, main text)
with the matrix element (10, main text).
As was first noted by Zakharov \cite{zakh65}, in the kinetic equation (11, main text)
in the isotropic case, the dispersion in $\omega _{k}$ at $%
k\rightarrow 0$ can be neglected, despite of a singularity due to the
product of two $\delta $-functions which after angle averaging turns out to
be integrable. As a result, the kinetic equation admits, except
thermodynamic distribution, a stationary power-law solution of the
Kolmogorov type: $n_{k}\propto $ $k^{-9/2}$, which corresponds to the ZS
spectrum $E(k)\propto k^{-3/2}$. Multiplying (11, main text) by $4\pi k^{3}\frac{%
	1}{(2\pi )^{3}}$ with account that angle average $\left\langle \delta \left(
\mathbf{k}-\mathbf{k}_{1}-\mathbf{k}_{2}\right) \right\rangle =2\pi
/(kk_{1}k_{2})$, we get
\begin{equation} \label{E(k)}
	\frac{\partial E(k)}{\partial t}=\frac{1}{32\pi ^{3}}k\int
	dk_{1}dk_{2}\left( W_{kk_{1}k_{2}}-W_{k_{1}kk_{2}}-W_{k_{2}kk_{1}}\right)
\end{equation}
where
\begin{equation*}
	W_{kk_{1}k_{2}}=k^{2}k_{1}^{2}k_{2}^{2}(n_{k_{1}}n_{k_{2}}-n_{k}n_{k_{2}}-n_{k}n_{k_{1}})\delta \left( k-k_{1}-k_{2}\right) .
\end{equation*}
Because of $\int E(k)dk=const$  the equation (\ref{E(k)}) can be rewritten
as a conservation law:
\begin{equation*}
	\frac{\partial E(k)}{\partial t}+\frac{\partial P}{\partial k}=0
\end{equation*}%
where $P$ is the energy flux which is found after integration of the
left-hand side of (\ref{E(k)}) over $k$
\begin{equation}
	\frac{\partial P}{\partial k}=-\frac{k}{32\pi ^{3}}\int dk_{1}dk_{2}\left(
	W_{kk_{1}k_{2}}-W_{k_{1}kk_{2}}-W_{k_{2}kk_{1}}\right) .  \label{derivative}
\end{equation}%
Let $n_{k}=Ak^{x}$ where $x$ is unknown exponent. It  gives for the quantity
$\frac{\partial P}{\partial k}$   a power-law function
\begin{equation}
	\frac{\partial P}{\partial k}=k^{y}I;\,\, y=2x+8  \label{flux-y}
\end{equation}%
where $I$ is a constant depending on $y$. Hence we have
\begin{equation}
	P=\frac{k^{y+1}}{y+1}I.  \label{flux}
\end{equation}%
Note, independence $P$ on $k$ takes place at $y=-1$ (where $I=0$ !) that
corresponds to the exact stationary solution of Eq. (11, main text) in the form
of the Zakharov-Sagdeev spectrum:
\begin{equation*}
	E(k)\propto k^{-3/2}.
\end{equation*}
In order to find dependence $I$ on $ x$ let us make Zakharov transformation
in the second and the third terms of (\ref{derivative}):%
\begin{eqnarray*}
	k_{1} &=&k\frac{k}{k^{\prime }},k=k^{\prime }\frac{k}{k^{\prime }}%
	,k_{2}=k^{\prime \prime }\frac{k}{k^{\prime }}; \\
	k_{2} &=&k\frac{k}{k^{\prime \prime }},k=k^{\prime \prime }\frac{k}{%
		k^{\prime \prime }},k_{1}=k^{\prime }\frac{k}{k^{\prime \prime }}.
\end{eqnarray*}%
This gives
\begin{equation} \label{flux}
	\frac{\partial P}{\partial k}=-\frac{A^{2}k}{32\pi ^{3}}\int
	dk_{1}dk_{2}k^{2}k_{1}^{2}k_{2}^{2}\left[ 1-\left( \frac{k}{k_{1}}\right)
	^{y}-\left( \frac{k}{k_{2}}\right) ^{y}\right] \left( kk_{1}k_{2}\right) ^{x}%
	\left[ k^{-x}-k_{1}^{-x}-k_{2}^{-x}\right] \delta \left(
	k-k_{1}-k_{2}\right) .
\end{equation}%
Note that stationary KZ spectrum takes place when
\begin{equation*}
	y=-1,\,\,  \mbox{or}\,\, x=-9/2.
\end{equation*}%
Then  integration of Eq. (\ref{flux}) relative to $k_{2}$ yields%
\begin{equation*}
	\frac{\partial P}{\partial k}=-\frac{A^{2}}{32\pi ^{3}}%
	k\int_{0}^{k}dk_{1}k^{2}k_{1}^{2}(k-k_{1})^{2}\left[ 1-\left( \frac{k}{k_{1}}%
	\right) ^{y}-\left( \frac{k}{k-k_{1}}\right) ^{y}\right] \left[
	kk_{1}(k-k_{1})\right] ^{x}\left[ k^{-x}-k_{1}^{-x}-(k-k_{1})^{-x}\right] .
\end{equation*}%
In terms of new variable $\xi =k_{1}/k$ ($0\leq \xi \leq 1$), this
expression takes the form:
\begin{equation*}
	\frac{\partial P}{\partial k}=-\frac{A^{2}}{32\pi ^{3}}k^{y}\int_{0}^{1}d\xi %
	\left[ 1-\left( \frac{1}{\xi }\right) ^{y}-\left( \frac{1}{1-\xi }\right)
	^{y}\right] \left[ \xi (1-\xi )\right] ^{x+2}\left[ 1-\xi ^{-x}-(1-\xi )^{-x}%
	\right]
\end{equation*}%
where
\begin{equation*}
	I(y)=-\frac{A^{2}}{32\pi ^{3}}\int_{0}^{1}d\xi \left[ 1-\left( \frac{1}{\xi }%
	\right) ^{y}-\left( \frac{1}{1-\xi }\right) ^{y}\right] \left[ \xi (1-\xi )%
	\right] ^{x+2}\left[ 1-\xi ^{-x}-(1-\xi )^{-x}\right]
\end{equation*}%
Now putting in (\ref{flux}) $y=-1+\alpha $ and considering the limit $%
\alpha \rightarrow 0$ result in the following expression for the energy flux
($P_{st}=\varepsilon $):
\begin{eqnarray}
	\varepsilon  &=&\frac{A^{2}}{32\pi ^{3}}\Lambda ,  \label{square} \\
	\Lambda  &=&\int_{0}^{1}d\xi \left[ \xi \ln \left( \frac{1}{\xi }\right)
	+(1-\xi )\ln \left( \frac{1}{1-\xi }\right) \right] \left[ \xi (1-\xi )%
	\right] ^{-5/2}\left[ 1-\xi ^{9/2}-(1-\xi )^{9/2}\right] .  \notag
\end{eqnarray}%
Here the convergent integral $\Lambda >0$ provides  the positive energy
flux, directed to the small scale region. Note also, that convergence of the integral means the locality of the spectrum.

Hence
\begin{equation*}
	A=4\pi \left( \frac{2\pi \varepsilon }{\Lambda }\right) ^{1/2}
\end{equation*}%
and respectively  the ZS spectrum takes the form
\begin{equation*}
	E(k)=\left( \frac{8\varepsilon }{\pi \Lambda }\right) ^{1/2}k^{-3/2}
\end{equation*}%
where the Kolmogorov-Zakharov constant is equal to%
\begin{equation*}
	C_{KZ}=\left( \frac{8}{\pi \Lambda }\right) ^{1/2}
\end{equation*}%
with $\Lambda =52.42$. Thus,  in the isotropic case we have $C_{KZ}\approx
0.22$.

\section{Numerical model and parameters}

Numerical integration of the system of equations (12, main text) and (13, main text) was carried
out in a cubic domain $(2\pi )^{3}$ with periodic boundary conditions over
all three coordinates. Time integration was performed using an explicit scheme
by means of the Runge-Kutta method of fourth order accuracy with step $%
dt=5\cdot 10^{-3}$. The equations were integrated over spatial coordinates
using pseudo-spectral methods with the total number of harmonics $%
N^{3}=512^{3}$. To calculate the fast Fourier transform, parallel computing
technology CUDA on graphics processors (GPU) was used. To suppress the
aliasing effect, we used a filter that nulls higher harmonics with a
wavenumber above $k_{a}\geq N/3$. The operator $\gamma _{k}$ responsible for
dissipation and the forcing term $\mathcal{F}(\mathbf{k},t)$ arising in (12, main text)
and (13, main text) are given in Fourier space as:
\begin{eqnarray*}
	\gamma _{k} &=&0,\quad k\leq k_{d}, \\
	\gamma _{k} &=&\gamma _{0}(k-k_d)^2,\quad k>k_{d}, \\
	\mathcal{F}(\mathbf{k},t) &=&F(k)\cdot \exp [iR(\mathbf{k},t)], \\
	F(k) &=&F_{0}\cdot \exp [-\lambda_0^4 (k-k_{1})^{4}],\quad k\leq k_{2}, \\
	F(k) &=&0,\quad k>k_{2}.
\end{eqnarray*}
Here $R(\mathbf{k,}t)$ are random numbers uniformly distributed in the
interval $[0,2\pi ]$, $\gamma _{0}$ and $F_{0}$ are constants.

The work presents three series of calculations for different regimes of
acoustic turbulence. The first calculation series is made to show the
possibility of realization of a weakly dispersive regime of weak acoustic
turbulence with small enough dispersion parameter, $a=2.5\cdot 10^{-3}$.
The second series presents pure non-dispersive regime with $a=0$ but keeping the
nonlinearity level small. Finally, the third calculation experiment is
carried out for non-dispersive wave regime but with sufficiently high level
of energy pumping to show the transition to a strongly nonlinear regime of
motion.

The dispersion length $a$ for the first experiment was chosen as $a=2.5\cdot
10^{-3}$. The maximum dispersion addition at the end of the inertial
interval, $k=k_{d}$, was $(k_{d}a)^{2}\approx 0.1$. Other parameters were: $%
k_{d}=110$, $k_{1}=3$, $k_2=6$, $\lambda_0=0.64$, $\gamma_{0}=10^{-1}$, $%
F_{0}=1.25\cdot 10^{-3}$. With this choice of parameters, the inertial
interval was more than one decade. For the second numerical experiment we
set, $a=0$, all other parameters except, $\gamma_0=10^{-2}$, were the same
as for the case of weak dispersion. The third series of calculations was
made for the pure non-dispersive regime but the quantity $F_0$ increased
from $1.25\cdot 10^{-3}$ to $40\cdot10^{-3}$, i.e., by more than an order of
magnitude. The selected value of $F_0$ is the maximal for our numerical
setup, since the width of the forming shocks becomes comparable with the
grid step. With such a large pumping amplitude, a ``bottleneck'' effect is
observed, which leads to wave energy accumulation near the dissipation
region $k_d$. To suppress the effect, we set $k_d=1$, i.e., the energy
dissipation occurs in all ranges of wavenumbers. Other parameters that
changed in this case were: $dt=2.5\cdot10^{-3}$, $\gamma_0=10^{-4}$.

\begin{figure}[h!]
	\centering
	\includegraphics[width=0.8\linewidth]{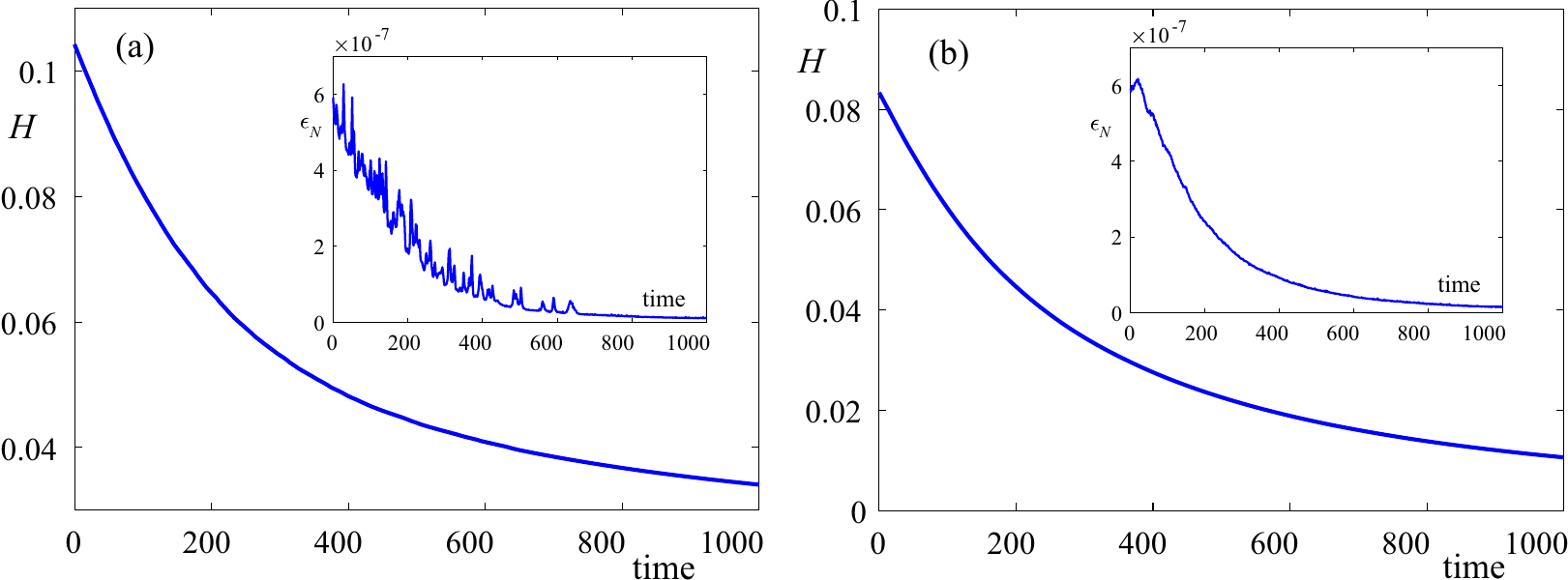}
	\caption{(Color online) The evolution of the Hamiltonian is shown for $%
		a=2.5\cdot 10^{-3}$ (a) and for $a=0$ (b). Insets show the evolution of the
		instantaneous energy dissipation flux. }
	\label{fig1}
\end{figure}

\begin{figure}[h!]
	\centering
	\includegraphics[width=0.8\linewidth]{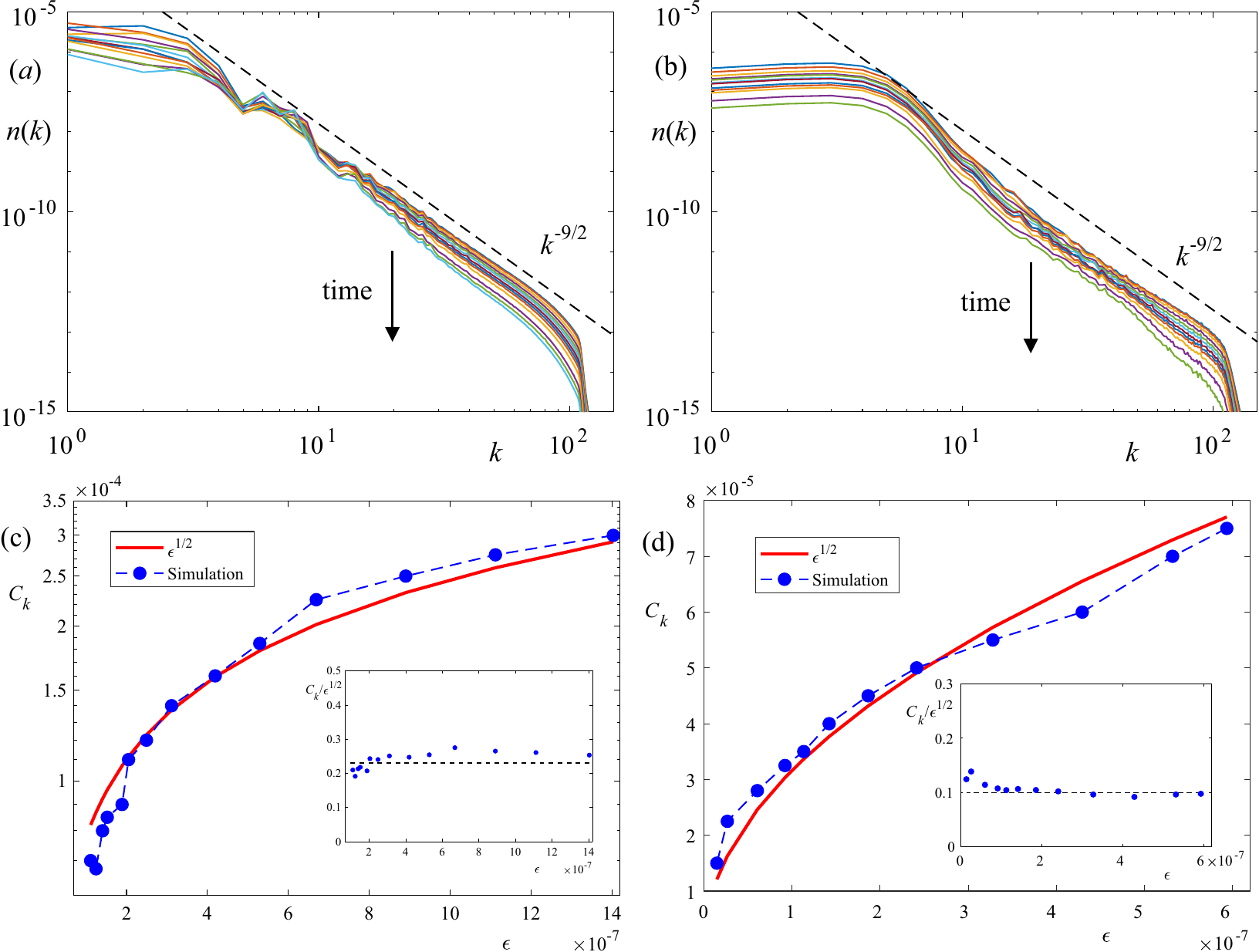}
	\caption{(Color online) The spatial turbulence spectra in terms of wave
		action $n(k)$ are shown in dependence on time for $a=2.5\cdot 10^{-3}$ (a)
		and for $a=0$ (b), black dashed line corresponds to KZ spectrum, $k^{-9/2}$.
		The coefficients $C_k$ are shown as a function of the measured energy
		dissipation flux are shown for $a=2.5\cdot 10^{-3}$ (c) and for $a=0$ (d).
		The red solid lines show the square root function $\protect\epsilon^{1/2}$.
		Insets show compensated values of $C_k/\protect\epsilon^{1/2}$ (KZ constant)}
	\label{fig2}
\end{figure}

\section{Estimation of the Kolmogorov-Zakharov constant}

In this section we present two series of calculations of weakly dispersive
and non-dispersive regimes of motion in order to calculate the KZ constants
for both cases. Numerical experiments simulate decaying turbulence for $F_0=0
$. The initial conditions were taken from the time instant, $t=2000$, from
Fig.~1~(a) and (b) corresponding to the quasi-stationary state. Calculations
were carried out over a time interval, $T=1000$. As can be seen from Fig. S
1, during this period of time the energy dissipates in several times, and
the dissipation flux almost stops, i.e., the system becomes almost linear.
The evolution of turbulence spectra in the decaying regime is shown in
Fig.~S~\ref{fig2}. It can be seen that the spectrum evolution is
self-similar with unchanged exponent $-9/2$, which corresponds to ZS
spectrum. To directly calculate the KZ constants, we measure the energy
spectrum constants, $E(k)=C_k k^{-3/2}$, in dependence on energy flux $%
\epsilon$. The measured dependencies are in well agreement with $%
\epsilon^{1/2}$, see Fig.~S~\ref{fig2}~(c) and (d) for both weakly
dispersive and non-dispersive regimes. The insets allow to directly estimate
the values of KZ constants for two regimes: $C_{KZ}\approx 0.24$ (weakly
dispersive) and $C_{KZ}\approx 0.1$ (non-dispersive).

\section{Analysis of an individual jet}

In this section, we present details of the analysis of an individual jet of
the function $u(\mathbf{k})$ in Fourier space. To analyze the effect of
divergence (wave diffraction), we consider a region in momentum space
occupied by a separate jet. We consider a set of rays (an individual jet)
passing through the origin of coordinates and extending along the $x$-axis,
see Fig.~2~(g)~and~(f) of the paper. As can be seen from the Fig.~2~(f), the rays
diverge by an angle in the order of $\pi/10$. Fig.~S~\ref{fig3}~(a) shows
the result of the filtration of initial spectrum by a cone with angle $\pi/10
$. One can actually see that all the rays fit inside the cone area. Fig.~S~%
\ref{fig3}~(b) demonstrates the inverse Fourier transform of the filtered
spectrum. It can be seen that the waves are not purely plane. The observed
waves are weakly modulated in the direction perpendicular to the wave
propagation. Applying the same filter to the function $\phi$, we calculate
the linear and nonlinear contributions of an individual jet to the
Hamiltonian (8). Direct calculations show that the contribution of linear
diffraction (wave divergence) turns out to be much greater than the energy
of nonlinear wave interaction.

\begin{figure}[h!]
	\centering
	\includegraphics[width=0.8\linewidth]{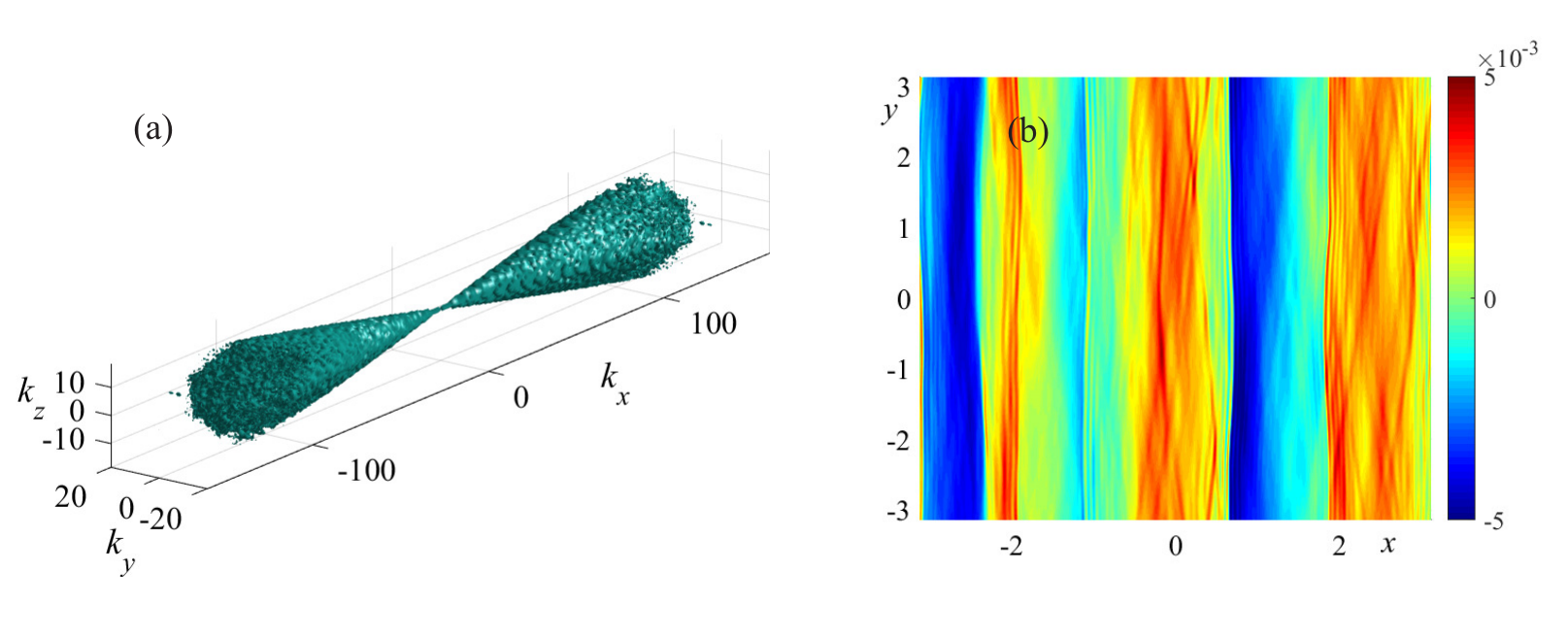}
	\caption{(Color online) Isosurface of the Fourier spectrum $|u(\mathbf{k}%
		)|=10^{-7}$ for the separate jet inside the angle $\protect\pi/10$. The
		inverse Fourier transform of the jet is shown in $\{x,y\}$ plane at the
		moment $t = 2000$ corresponding to the quasi-stationary state.}
	\label{fig3}
\end{figure}

\section{''Bottleneck'' effect}

\begin{figure}[h!]
	\centering
	\includegraphics[width=0.8\linewidth]{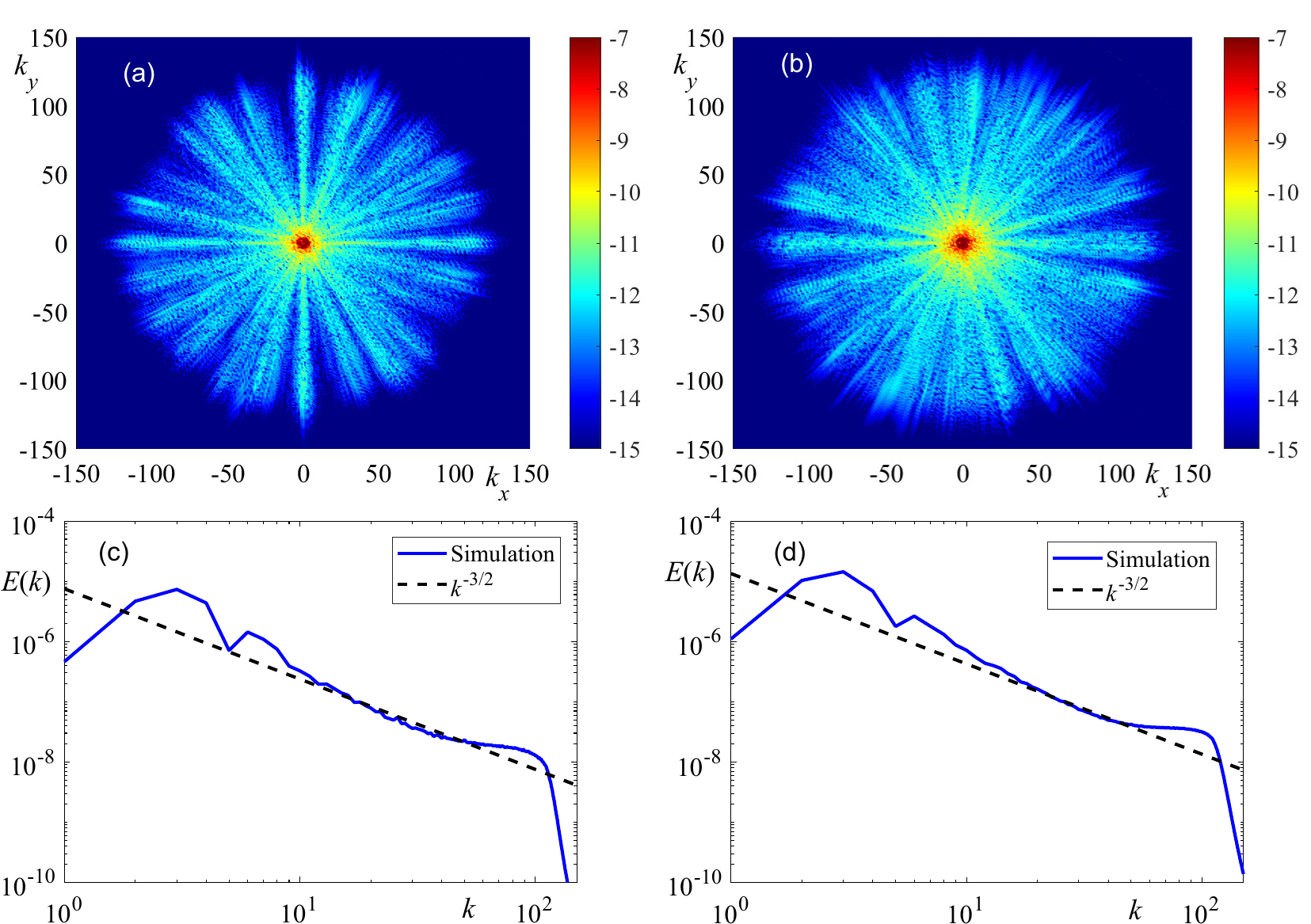}
	\caption{(Color online) (a) and (b) Fourier spectra of $u(\mathbf{r})$ in $%
		\{k_x, k_y\}$ plane. (c) and (d) The spatial turbulence spectrum in terms of
		spectral energy density. (a), (c) and (b), (d) correspond to $F_0=2.5\cdot
		10^{-3}$ and $F_0=5\cdot 10^{-3}$, respectively.}
	\label{fig4}
\end{figure}

In this section we provide details of the analysis of the bottleneck effect.
With this effect, energy accumulates in a region close to the dissipation
scale $k_d$. The effect is observed with increasing amplitude of energy
pumping. Fig.~S~\ref{fig4} shows the Fourier spectra of the function $u(%
\mathbf{r})$ in the quasi-stationary regime for $F_0=2.5\cdot 10^{-3}$ and $%
F_0=5\cdot 10^{-3}$. It can be seen that as the energy pumping level
increases, the beams in Fourier space become broader. With a further
increase in the level of nonlinearity, rays from neighboring cones begin to
intersect. At the same time, a peak near $k_d$ appears in the turbulence
spectra, see Fig.~S~\ref{fig4}~(c) and (d). To minimize the bottleneck
effect, we introduce dissipation over the entire wavenumber range, i.e., $%
k_d=1$. Thus, at a large pumping amplitude, the distribution of waves in
Fourier space become quasi-isotropic, and the turbulence spectrum acquires
shock-wave character, i.e., $E(k)\propto k^{-2}$.

\section{Characteristic nonlinear time comparison}

\begin{figure}[h!]
	\centering
	\includegraphics[width=0.5\linewidth]{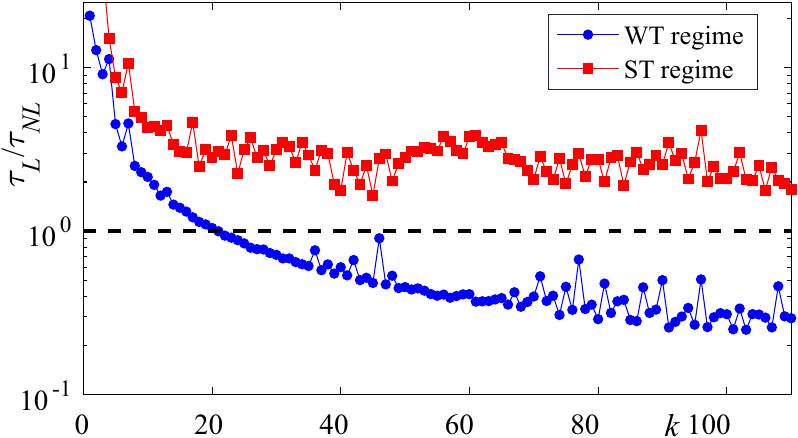}
	\caption{ Linear to nonlinear time ratios for $a=0$. }
	\label{figs5}
\end{figure}

Fig.~3~(d) shows the space-time Fourier transform of the function $%
u(\mathbf{r},t)$. One can indeed see that frequency broadening $%
\delta_{\omega}$ is much higher in comparison with the weakly turbulent
regime of motion shown in Fig~2~(d)~and~(h). The given space-time Fourier spectra allow to estimate the
frequency broadening $\delta_{\omega}$ (nonlinearity level) for each $k$
\cite{Naz22}:
\begin{equation*}
	\delta_{\omega}=\left[\int_0^{\infty}(\omega-\omega_k)^2|u( k,\omega)|^2
	d\omega/\int_0^{\infty}|u(k,\omega)|^2 d\omega\right]^{1/2},
\end{equation*}
with $\omega_k=|\mathbf{k}|$. In fact, the quantity $\delta_{\omega}$
determines the characteristic nonlinear time $\tau_{NL}=1/\delta_{\omega}$.
The applicability criterion for weak turbulence theory \cite{KZ-book} can be
written by the following way: $\tau_L/\tau_{NL}\ll 1$, where $\tau_L$ is the
characteristic linear time, in our case $\tau_L=[\Omega_0^2k/2]^{-1}$. Fig.~S~\ref{figs5} shows the calculated time relation for weakly (WT) and strongly
(ST) turbulent regimes. One can see that the difference between the
parameters is near an order of magnitude. The value of $\tau_L/\tau_{NL}$
for ST regime is above the threshold of weak turbulence applicability,
whereas the measured time relation for WT regime is at least in five times
less than unity, i.e., the criterion is satisfied in the weakly nonlinear
case.

%\twocolumn


\begin{thebibliography}{99}
\bibitem{KZ-book} V.E. Zakharov, V.S. L'vov, G. Falkovich, Kolmogorov
Spectra of Turbulence I: Wave Turbulence / Berlin Springer-Verlag, 1992.

\bibitem{naz-book} S. Nazarenko, Wave Turbulence (Springer Berlin
Heidelberg, series: Lecture Notes in Physics, 2011).

\bibitem{FM} E. Falcon, N. Mordant, Annu. Rev. Fluid Mech. \textbf{54}, 1
(2022).

%\bibitem{Gurevich} V.L. Gurevich, Kinetika fononnykh sistem (Phonon System
%Kinetics) Nauka. Moscow, 1980 [in Russian].

%\bibitem{Galtier} S. Galtier, J. Fluid Mech., \textbf{974}, A24 (2023).


%\bibitem{grav} M. Hindmarsh, et al. Phys. Rev. Lett. \textbf{112} (4),
%041301 (2014).

%\bibitem{Shuryak} T. Kalaydzhyan, E. Shuryak, Phys. Rev. D, \textbf{91}%
%(8), 083502 (2015).

\bibitem{zs-70} V.E. Zakharov, R.Z. Sagdeev, Sov. Phys. Dokl. \textbf{15},
439 (1970).

\bibitem{Newell} A. C. Newell and P. J. Aucoin, J. Fluid Mech. \textbf{49},
593 (1971).

\bibitem{AT} V. S. L'vov, Y. L'vov, A. C. Newell, and V. Zakharov, Phys.
Rev. E \textbf{56}, 390 (1997).

\bibitem{zakh65} V. E. Zakharov, J. App. Mech. Tech. Phys., \textbf{6}(4),
22 (1965).

\bibitem{Naz22} A. Griffin, G. Krstulovic, V. S. Lvov, S. Nazarenko, Phys.
Rev. Lett. \textbf{128}, 224501 (2022).

%\bibitem{KuznetsovTuritsyn1982} E.A. Kuznetsov, S.K. Turitsyn, Sov. Phys.
%JETP \textbf{55}, 844 (1982).

%\bibitem{KuznetsovMusher} E.A. Kuznetsov, S.L. Musher, Sov. Phys. JETP
%\textbf{64}, 947 (1986).


\bibitem{ZakharovKuznetsov2012} V. E. Zakharov, E.A. Kuznetsov, 	Phys. Uspekhi \textbf{55}, 535-556 (2012).


%\bibitem{ZakharovKuznetsov1997} V.E. Zakharov, E.A. Kuznetsov, Phys.
%Uspekhi, \textbf{40}, 1087-1116 (1997).

\bibitem{KP} B.B. Kadomtsev, V.I. Petviashvili, Dokl. Akad. Nauk SSSR
\textbf{208}, 794 (1973).

\bibitem{Kuz04} E. A. Kuznetsov, JETP Lett. \textbf{80}, 83 (2004).

\bibitem{Burgers} J. M. Burgers, Adv. Appl. Mech. \textbf{1}, 171 (1948).

\bibitem{KochurinKuznetsov} E.A. Kochurin, E.A. Kuznetsov, JETP Lett.
\textbf{116}, 863 (2022).

\bibitem{SP} See Supplemental Material at https://

\bibitem{landau} L.D. Landau and E. M. Lifshitz, Fluid Mechanics, p. 135
(Oxford: Pergamon Press, 1987).

\bibitem{kin1} C. Connaughton, P.L. Krapivsky, Phys. Rev. E \textbf{81},
035303(R) (2010).

\bibitem{kin2} W. Walton, T. Minh-Binh, SIAM J. Sci. Comput. \textbf{45}
(3), B467 (2023)

%

\end{thebibliography}
\end{document}